# Inductive decision based Real Time Occupancy detector in University Buildings.

Nikita Jain
Research Scholar, USICT, GGSIPU
Department of Computer Science
Bhagwan Parshuram Institute of Technology

Rachita Gupta

Bhagwan Parshuram Institute of Technology
New Delhi, India

*Abstract*—**The ability to estimate College Campus Occupancy for Classrooms and Labs in real time has become one of the major concerns for various Academicians, authorities and administrators ,where still a manual attendance marking system is being followed. Using a low budget multiple sensor setup installed in a college auditorium, the goal is to build a real-time occupancy detector. This paper presents an Inductive real time Decision tree based classifier using multiple sensor dataset to detect occupancy. Using simple feature based thresholds, Reverberation time which comes out to be a novel as well as most distinguishing feature sampled at various frequencies over a given time interval was used to detect the occupancy with an accuracy of %.Addition of various other sensor data, decreased the accuracy of classification results. The detector setup can be used in various college buildings to provide real time centralised occupancy status thus automating the manual attendance system being used.**

## I. INTRODUCTION

Knowing how many students occupy a classroom, labs or conference hall is a key component of building an automated occupancy detection system. University Buildings with huge classrooms and labs often employ a manual attendance system. Although, the current occupancy detection via manual counting serves as a highly accurate measure of counting the number of students in a room, however it accounts for a high manual and maintenance based human effort.

This paper describes results from a project to design, implement, Validate and propose an Inductive Decision based Framework to detect occupancy in University/College campus's Classrooms and Conference Halls.

Here we implement better sensing for accurate results rather than large dataset achieved by combining the given approach:-

- *Installing a low cost distributed sensor network*
  To improve the accuracy of results a distributed multi sensor network was installed in a college building's classroom.
- *Proposing a new feature based Analysis Algorithm for processing the collected data set.*

Our approach for detecting the occupancy is based on inductive based decision tree algorithm where prediction and diagnosis can be done even in noisy and incomplete information dataset.

Inductive decision tree (ID3) can be seen as a divide-and-conquer for object classification. Decision tree presents a procedure for determining the class of given instance where a training instance is taken as a list of attribute-value pairs. Here each instance is assumed to be either positive instances i.e *target concept* or negative instances [1].

In this paper we propose a decision based tree algorithm for multiple sensor information gained in real time from a university classroom under observation. In respect to an object taken as one training instance following are the attributes studied for each instance:

- Temperature
- $CO_2$
- Frequency

Using every parameter mentioned above we plan to implement the suited based decision tree to detect occupancy.

In our current research, we present a new parameter *Reverberation time* which is time in seconds for a sound to decrease in value by 60 dB [2]. An echo is produced by sound waves bouncing, or reverberating, from one or more hard surfaces and which may last for several seconds. A room that has a long reverberation time sounds noisy, lively and thus occupied. Using the above parameters, we hereby present a novel multi sensor occupancy detector based on realtime sensing.

This paper is structured as follows: Section II explores the past and ongoing research for occupancy detection in real time Section III presents the experimental setup used for detecting occupancy in university classrooms. Consequently in section IV, methodology used to detect occupancy based on parameters sensed is discussed .In Section IV we present, the results and analysis followed by Section V that concludes the paper.

## II. LITERATURE REVIEW

Occupancy detection has been a much searched and discussed topic among researchers in past.. Because of it's practical applications, sensor network as a method to detect occupancy is being formulated and documented.

Authors in [3] have presented a detailed study of how CO2 can be used as a driver for detecting occupancy in office and residential Buildings using stochastic models. The authors here have provided a new area of research where camera based video streaming is being used to detect occupancy.

Similarly, In [4] belief based networks is being used to detect occupancy which falls under the category of probabilistic models. Here, authors have used infrared occupancy sensors with Bayesian network model to predict occupancy realtime.

Various occupancy detector schemes and methodologies have been surveyed well in [5].Here demand driven control applications have been surveyed. It clearly states how thermal, visual , air quality control systems which account for 30% of energy consumed in office building are being hampered due to inability of accurate occupancy detector. Authors have categorized the occupancy systems as:

- CO2 based
- Passive Infrared
- Ultrasonic detection
- Sound detection
- Computer activity

Apart from above systems, a 24 hour image based detection occupancy detector has been built,validated and presented in [6].Here Authors have used SVM based classifier

Treating Occupancy detection as a classification problem and quite imperative to today's commercial building applications, We hereby:

- Present a low budget multiple sensor distributed network
- The sensor apparatus consists of frequency sensors used to further calculate Reverberation time: A measure to estimate number of echoes in a room
- Using various parameters in conjunction with Reverberation time occupancy is detected in realtime.

## III. APPARATUS:ROOM DETECTOR

The room used here under observation was taken from a known college building. This room is actively used for conferences, seminar and practical sessions on daily basis. Here, the room configuration is
- Length- 70 feet
- Breadth- 30 feet and
- Height- 12 feet

The parameters mentioned above have been demonstrated crucial to occupancy detection [7]. To employ a multi parameter sensor network, a distributed network sensing various parameters was deployed in the room under observation. Data was collected over a period of 7 days in the campus itself. The multi sensor description with make model and quantity deployed in the test area is given below in Fig 1

| Type | Make | Model | Quantity | Price |
|---|---|---|---|---|
| **CO2** | mhestore2009 | MHZ-19 | 2 | 1320 |
| **Temperature** | robodo.electronics | DS18B20 | 1 | 145 |
| **Frequency** | Researchdesignlab | RDL/VBS/13/001/V1.0 | 2 | 395 |

**Fig 1 :** Sensors used in College classroom study

A physical configuration setup if the test room is given in Figure 2.
As discussed in previous section, reverberation time is a measure to detect noise in a room. Given a frequency sensed by the sensors, reverberation time(T) can be calculated using equation:

$$T = 0.161 * V / S * α` \qquad (1)$$

Where: V is Volume of the room
S is Surface area of the room
and  α` is Mean Absorption Coefficient

Mean absorption coefficient, α, is found from the area and absorption coefficient for each surface of the enclosing space. All the absorbing surfaces within the space, such as seats and people in a theatre, are included in the overall sound absorbing ability of the room [2].

$$α` = A1 × α1 + A2 × α2 + A3 × α3 / A1 + A2 + A3 \qquad (2)$$

where: A1,A2,A3…. is surface area of surface number 1,number 2 …respectively and  α1, α2…. is absorption coefficient of  surface number 1,number 2 ….

As every material absorb different amount of sound at different frequency therefore: Using the above formula (2) and physical configuration of the test room, Absorption coefficient of the 3 types of material used in the room under observation were sensed at different frequencies and are hence presented in Table 1 . Considering the calculated absorption coefficient also sensed at different frequencies and reverberation time (T) is calculated for every different frequency. It is important to cap the threshold for reverberation time which is one of the prime feature being used by our detector setup.

A higher value of T denotes noise in the room *in terms of* echoes and hence justifies no occupancy in room. Similarly a low value of T denotes there is no echo therefore justifying an occupied room. Clearly, room resonance or frequency is affected by the range of sound or noise present[8]. Using these frequency values sensed by the sensors deployed and formula (2) respective absorption coefficient is calculated for all materials that are used in the given specific room.

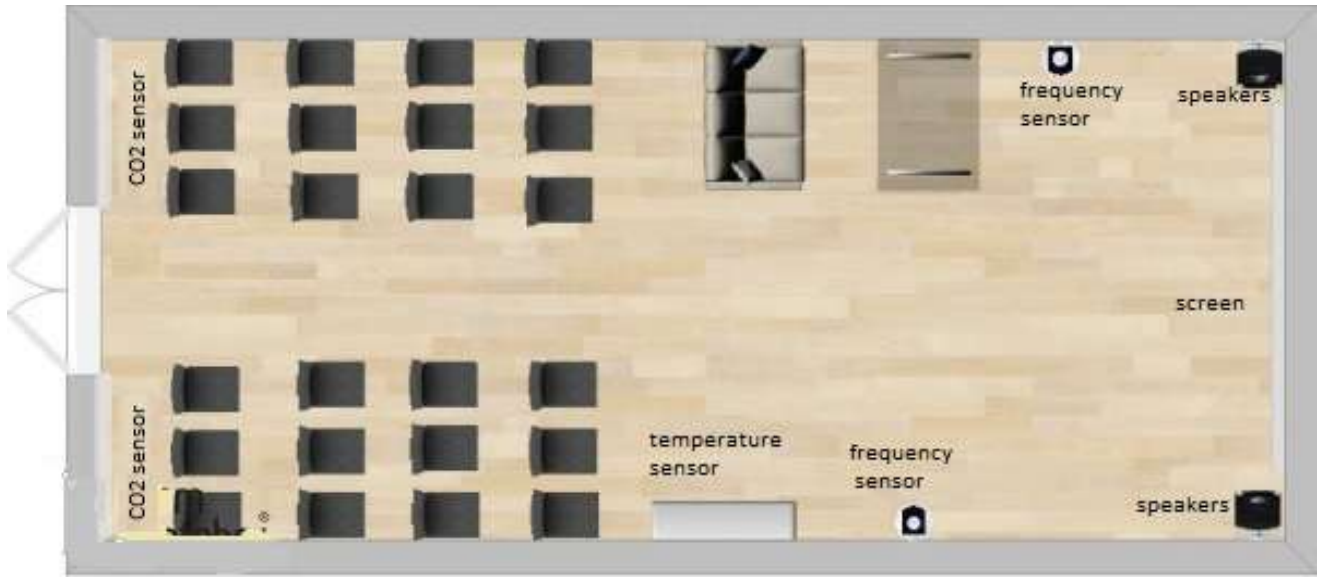

Fig 2: Physical Configuration of Occupancy Detector setup with multiple sensors installed in a College Conference hall.

## IV. INDUCTIVE DECISION IMPLEMENTATION: METHODOLOGY

Researchers in past have demonstrated that decision tree algorithm is best suited for dual classification problem [9]

Figure 3 presents the best suited decision tree selected on the basis of the prepared dataset given to it. As shown the selected tree detects occupancy using all three features Temperature, CO2 and reverberation time as calculated above.

In this paper, we have used an open source application KNIME (Konstanz Information Miner) 2.1.2.0024559 [10] as the environment to perform the data analysis.

All the sensed values were sample over a period of 7 days with a time interval of about 50 minutes i.e a standard lecture time in a college routine days. The values of these parameters were taken at the start of every lecture which ensured the known occupancy taken here as 1.

| Frequency | Absorption coefficient | | |
|---|---|---|---|
| | Brick | Concrete | Carpet |
| 3087 | 0.020448859 | 0.05 | 0.612724866 |
| 2922 | 0.020448859 | 0.05 | 0.612724866 |
| 3029 | 0.029637413 | 0.05 | 0.623 |
| 235 | 0.047578626 | 0.02 | 0.480870573 |
| 1243 | 0.03 | 0.047578626 | 0.612724866 |
| 2750 | 0.028228353 | 0.05 | 0.608789159 |
| 2017 | 0.028228353 | 0.05 | 0.612 |
| 562 | 0.030277425 | 0.024392294 | 0.5 |
| 1003 | 0.03 | 0.042393423 | 0.531691993 |
| 2409 | 0.023168347 | 0.05 | 0.612724866 |
| | | | |

**Table 1** Absorption Coefficients of room materials at different Frequencies.

*Decision Tree Method*

Decision tree as the name suggests is used for classification by traversing the tree of all permutations of decisions. The internal node here uses the feature values here the sensed parameters with the threshold values and thus a decision is taken regarding occupancy in binary at the child nodes.

The tree selected here is trained for predicting occupancy as **0 :for not occupied** and **1: for occupied.**

Inductive based decision tree treats every dataset entry as a single object with its attribute values as nodes of the tree. Using the Information gain theory, root of the tree is selected. Out of the given parameters nodes are only created if they reach K points in the given dataset.

Here we had 6 sensors, 3 types of feature were sampled over 8 slots of time widths over 7 days .This created a vector of size 1008 feature values sampled over a 50 minute slot

It is important to note that under permutation of reverberation time being included in prediction followed by Temperature accounts for the highest accuracy.

Clearly not including reverberation time lowers the accuracy the most .As no signs of major change in accuracy is observed when $CO_2$ as a feature is included. Hence $CO_2$ as a decision making parameter is less crucial for occupancy detection.

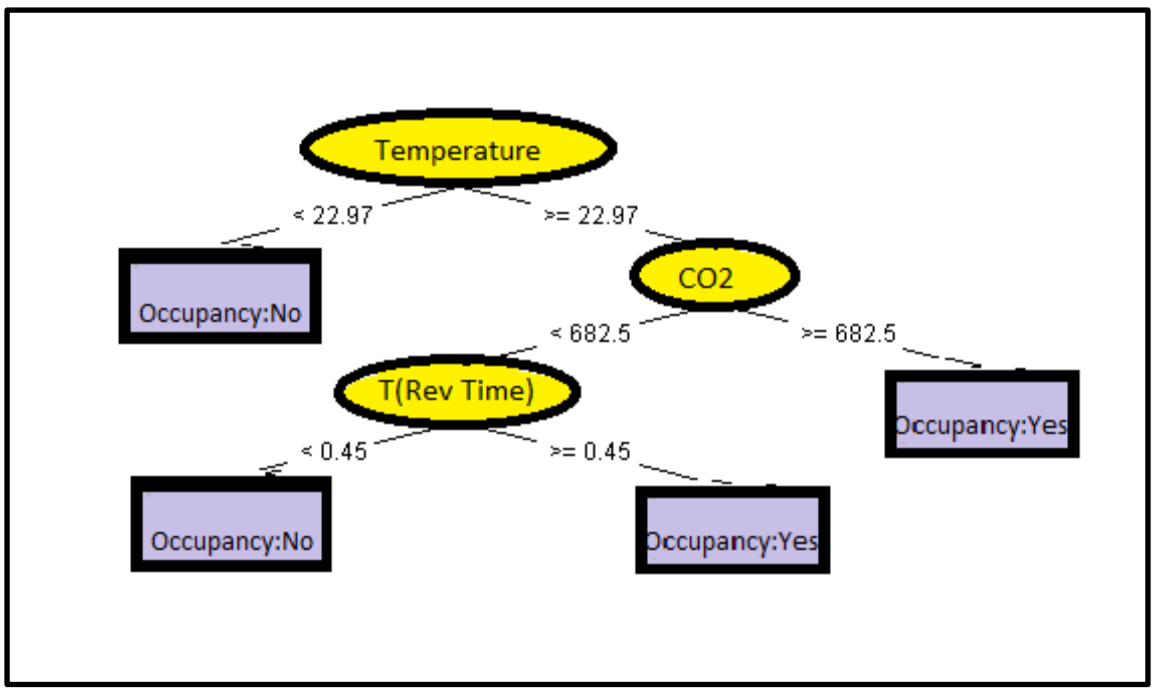

Fig. 3 Occupancy Detection Decision Tree based on multiple sensed parameters.

## V. RESULTS

Our experiment consisted of various trials taking every permutation of features selected.

The mentioned permutations were then evaluated under 7 fold cross validations. This accounted for 6 day data to be used for validation where 1 day data was used for training the decision classifier .A result of each of the 6 day data where occupancy was predicted by the classifier is presented in Table 2.

Based on above discussion a combined methodology for occupancy detection in figure 5 is proposed where reverberation time as a crucial parameter is taken.

The methodology proposed clearly takes a threshold level of 0.45 seconds for detecting occupancy Using the proposed methodology, 6 day data was fed into classifier to predict occupancy. The predicted occupancy was compared with known occupancy to calculate the accuracy per day.

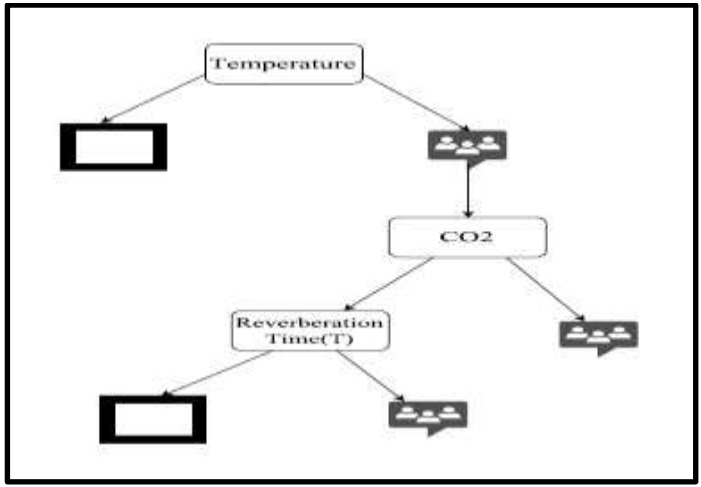

| CO2 | Temperature | Reverberation time | Accuracy |
|---|---|---|---|
| **Included** | Not Included | Included | 97.890 |
| **Included** | Included | Not included | 96.010 |
| **Included** | Included | Included | 97.853 |
| **Not Included** | Included | Included | 98.213 |
| **Not Included** | Not Included | Included | 98.111 |
| **Not Included** | Included | Not Included | 97.554 |

Fig 4: Top level occupancy detection decision tree for multiple sensed parameters

Table 2: Top level occupancy detection decision tree for multiple sensed parameters

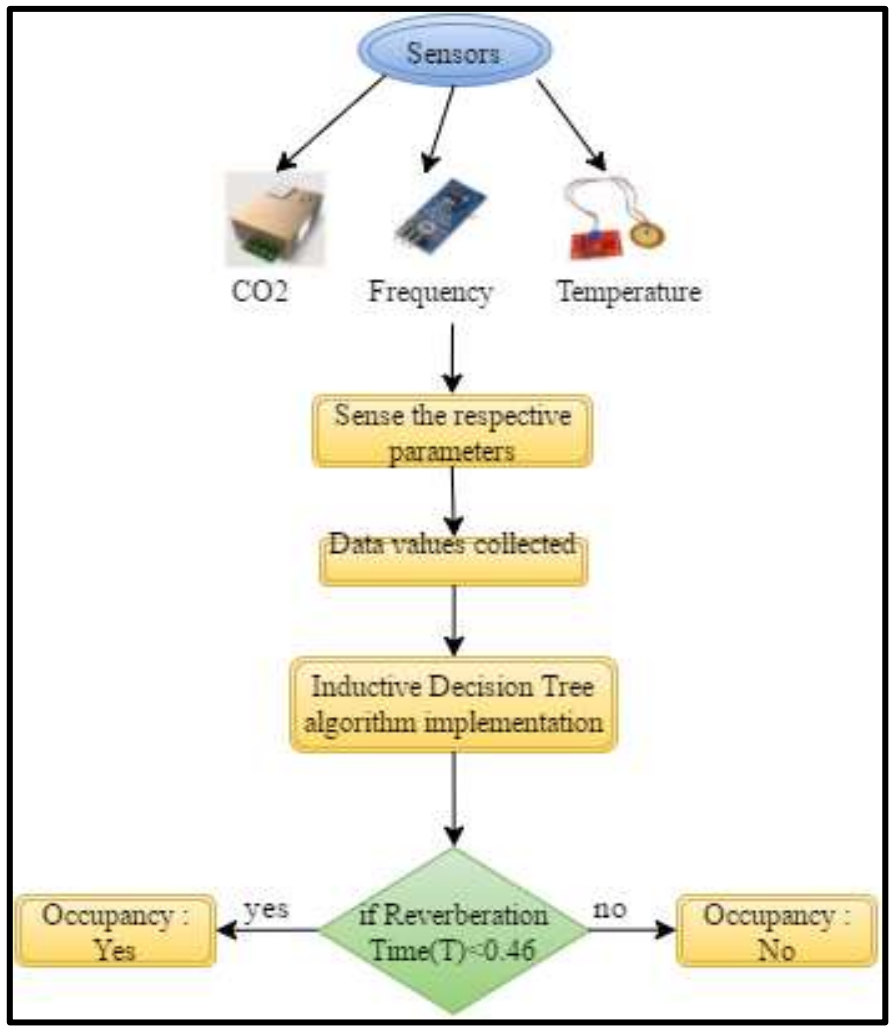

Fig 5: Occupancy detection methodology

*Aspect Value Analysis*

Table 3.presents the aspect evaluation done on the raw sensed data to further train classifier and generate occupancy alert. Table 4 shows the predicted occupancy by the classifier for some dataset values. The values in difference were counted and a mean accuracy was calculated as given in Table 2.

**Table 1** Multiple features sensed across the observation room with known Occupancy

| Temperature | CO2 | Reverberation Time(T) | Occupancy |
|---|---|---|---|
| 23.18 | 721.25 | 1.459188744 | 0 |
| 23.15 | 714 | 1.456123701 | 0 |
| 23.15 | 713.5 | 1.473628013 | 0 |
| 23.15 | 708.25 | 1.635583564 | 0 |
| 23.1 | 704.5 | 0.46944317 | 1 |
| 23 | 681.5 | 0.451661291 | 1 |
| 22.945 | 685 | 0.460310371 | 1 |
| 22.945 | 685 | 0.520755814 | 1 |
| 22.89 | 689 | 0.462277467 | 0 |
| 22.89 | 689.5 | 0.456447871 | 1 |

| Temperature | CO2 | Reverberation Time(T) | Occupancy |
|---|---|---|---|
| 23.18 | 721.25 | 1.459188744 | 0 |
| 23.15 | 714 | 1.456123701 | 0 |
| 23.15 | 713.5 | 1.473628013 | 1 |
| 23.15 | 708.25 | 1.635583564 | 0 |
| 23.1 | 704.5 | 0.46944317 | 1 |
| 23 | 681.5 | 0.451661291 | 0 |
| 22.945 | 685 | 0.460310371 | 1 |
| 22.945 | 685 | 0.520755814 | 0 |
| 22.89 | 689 | 0.462277467 | 0 |
| 22.89 | 689.5 | 0.456447871 | 1 |

**Table 4** Multiple features sensed across the observation room with predicted occupancy

V. CONCLUSION

Occupancy prediction is imperative as far as less technically centralized building are concerned .College campus building account in one of them where still manual attendance marking system is being followed in majority of colleges.This paper proposes a low budget multisensory based decision driven methodology to predict Occupancy in realtime. Though it is intuitive to use CO2 and temperature as crucial parameters, our study proposes a new parameter as reverberation time which calculates the amount of echoes being produced in a room at different level of sampled frequency. Using this new proposed feature with inductive decision based approach,an accuracy of approximately 98.213% is achieved in predicting occupancy in realtime.This low budget multiple sensor distributed network can be used under centralized installation to detect lab's, classroom, conference hall occupancy in any respective university or college campus.   References

[1] Quinlan J.R 1986.Induction of Decision Trees.Journal of Machine learning 81-106 Volume 1

[2]http://ebooks.narotama.ac.id/files/Building%20Services%20Engineering%20(5th%20Edition)/Chapter%2014%20Room%20Acoustics.pdf

[3] Davide Calì*, Peter Matthes, Kristian Huchtemann, Rita Streblow, Dirk Müller . CO2 based occupancy detection algorithm: Experimental analysis

and validation for office and residential buildings. Elsevier journal,Building and Environment 86 (2015) 39e49

[4] Robert H. Dodier , Gregor P. Henze , Dale K. Tiller ,Xin Guo Building occupancy detection through sensor belief networks.Elsevier journal,Energy and BUILDINGS38 (2006) 1033–1043

[5] Timilehin Labeodan, Wim Zeiler, Gert Boxem, Yang Zhao. Occupancy measurement in commercial office buildings for demand-driven control applications—A survey and detection system evaluation.. Elsevier Journal, Energy and Buildings 93 (2015) 303–314

[6] Huang-Chia Shih. A robust occupancy detection and tracking algorithm for the automatic monitoring and commissioning of a building. Elsevier Energy and Buildings 77 (2014) 270–280

[7] Real-Time Occupancy Detection using Decision Trees with Multiple Sensor Types.

[8] https://en.wikipedia.org/wiki/Room_acoustics

[9] A. N. Meenakshi, B. I. Juvanna, Implementation of decision tree algorithm c4.5 - International Journal of Scientific and Research Publications, Volume 3, Issue 10, October 2013 1 ISSN 2250-3153

[10] BERTHOLD, M. R., CEBRON, N., DILL, F., GABRIEL,
T. R., K¨OTTER, T., MEINL, T., OHL, P., SIEB, C., THIEL,
K. AND WISWEDEL, B., 2007. KNIME: The Konstanz Information Miner. In Studies in  Classification, Data Analysis, and Knowledge Organization (GfKL 2007). Springer.